\documentclass[twocolumn,showpacs,amsmath,amssymb,prl,showkeys,floatfix]{revtex4}

\usepackage{graphicx}
\usepackage{dcolumn}
\usepackage{bm}

\begin{document}

\preprint{APS/123-QED}

\title{Surface contribution to the superconducting properties
of MgB$_{2}$ single crystals}

\author{A. Rydh}
 \altaffiliation[Also at ]{Solid State Physics, Royal Institute of 
 Technology (KTH), Stockholm-Kista, Sweden}
 \email{rydh@anl.gov}
\author{U. Welp}
\author{J. M. Hiller}
\author{A. Koshelev}
\author{W. K. Kwok}
\author{G. W. Crabtree}
\affiliation{%
Materials Science Division, Argonne National Laboratory, 9700 S. Cass 
Ave., Argonne, IL 60439, USA}
\author{K. H. P. Kim}
\author{C. U. Jung}
\author{H.-S. Lee}
\author{B. Kang}
\author{S.-I. Lee}
\affiliation{%
NCRICS and Dept. of Physics, Pohang University of Science and 
Technology, Pohang 790-784, Republic of Korea}

\date{\today}

\begin{abstract}
We demonstrate direct evidence of possible surface superconductivity
on small, well-shaped MgB$_{2}$ single crystals.  Transport
measurements in the range $H < 1.6 H_{c2}^{c}$, where $H_{c2}^{c}$ is
the bulk upper critical field for the $c$ axis, show non-Ohmic and
strongly angular dependent resistivity.  Studies of the alignment of
$\mathbf{H}$ with selected crystal surfaces, transport and specific
heat measurements on the same crystal, and a physical sculpturing of
the crystal surfaces using a focused ion beam all support the
conclusion.  Similar, albeit less pronounced results are obtained for
fields in the basal plane.
\end{abstract}


\maketitle

The binary MgB$_{2}$ compound has been subject of intense studies
since the discovery of its superconducting properties at temperatures
up to 39 K \cite{nagamatsu01}.  The superconductor has been found to
be of phonon-mediated BCS type but with a multitude of novel
properties, mainly arising from its complex, two-band ($\pi$ and
$\sigma$) Fermi surface \cite{Choi02,Liu01}.  In particular, MgB$_{2}$
is the first example of a superconductor showing two distinct
superconducting gaps \cite{Bouquet01}.  The two-gap nature is revealed
in the macroscopic properties of MgB$_{2}$ through a pronounced
temperature dependence of the anisotropy of the upper critical field
\cite{Angst02,Lyard02,Welp03,Sologubenko02}.  The effect of the
two-gap structure on vortex dynamics is presently under intense
investigation.  Although STM \cite{Eskildsen02} and specific heat
measurements \cite{Bouquet02} indicate a rapid suppression of the
$\pi$ gap with increasing magnetic field its effect on
magneto-transport properties remains controversial.  In particular, a
pronounced broadening of the resistive transitions in magnetic fields
applied along the $c$ axis has been observed.  Whether this broadening
is related to the two superconducting gaps \cite{Pradhan02} or other
phenomena such as vortex dissipation and vortex lattice melting
\cite{Eltsev02}, surface barriers \cite{MasuiCond}, superconducting
fluctuations \cite{MasuiCond}, or surface superconductivity
\cite{Lyard02,Welp03,Sologubenko02} remains to be settled.

In this Letter we present definitive indications of a surface
superconducting state in well-shaped MgB$_{2}$ single crystals.  From
angular dependent transport and specific heat measurements, we
demonstrate that the broadening of the resistive transitions for
$\mathbf{H} \parallel \mathbf{c}$ as well as for $\mathbf{H} \parallel
\mathbf{ab}$ occurs \emph{above} the bulk upper critical field, while
there is virtually no broadening of the transitions in the mixed
state.  The result is confirmed by directly modifying the crystal
surface geometry using a focused ion beam (FIB).

MgB$_{2}$ crystals with typical maximum dimensions of 50~$\mu$m were
obtained through a high pressure heat treatment of a mixture of Mg and
B in excess Mg \cite{Jung02}.  The resulting crystals were well shaped
with smooth, hexagonal facets.  Scanning electron microscopy (SEM) and
high-resolution transmission electron microscopy confirmed the absence
of grain boundaries or other correlated defects \cite{Kim02}.  The
crystals had $T_{c} \approx 36$~K, $\Delta T_{c} \approx 0.15$~K
(at low current), and a bulk $H_{c2}^{c}(0) \approx 3.5$~T. Transport
measurements were performed using standard DC and AC techniques.  A
total of five crystals were studied, all showing similar behavior with
slight variations in the exact shape of the resistive transition. 
We present results on two of these crystals.

\begin{figure}
\includegraphics[width=1.0\linewidth]{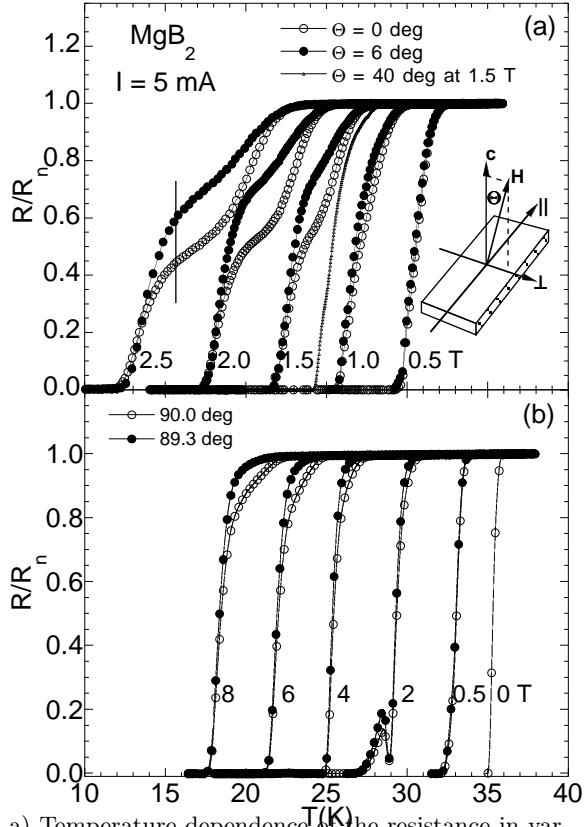}
\caption{a) Temperature dependence of the resistance in various fields
directed along and at 6$^{\circ}$ off the $c$ axis.  At 1.5~T data for
an angle of 40$^{\circ}$ are also included.  The inset defines the
geometry.  b) Similar data for fields applied around the $ab$ plane. 
For clarity only 2\/\% of the data points are displayed.}
\label{Fig1}
\end{figure}

Figures~\ref{Fig1}a and \ref{Fig1}b show the resistive transitions
in various fields around the $c$ axis and around the basal ($ab$) plane,
respectively.
The transitions broaden significantly with increasing field around $c$
as reported earlier
\cite{Welp03,Sologubenko02,Pradhan02,Eltsev02,MasuiCond}.  This
broadening evolves in a characteristic two stage fashion with a
gradual onset at $T_{\mathrm{on}}$ and a steep drop at lower
temperature, which we are going to identify with $T_{c2}(H)$.  The
gradual onset broadens in increasing field, while the steep drop at
$T_{c2}(H)$ stays sharp.  In addition to a strong current dependence
the broadening is also strongly angular dependent.  As the applied
field is tilted away from the $c$ axis the gradual resistive onset is
rapidly suppressed and the steep resistive drop becomes more
pronounced.  The gradual onset is largely eliminated at angles around
$40^{\circ}$ (in 1.5~T).  Similar behavior is seen for fields applied
along the $ab$ plane (Fig.~\ref{Fig1}b).

\begin{figure}
\includegraphics[width=1.0\linewidth]{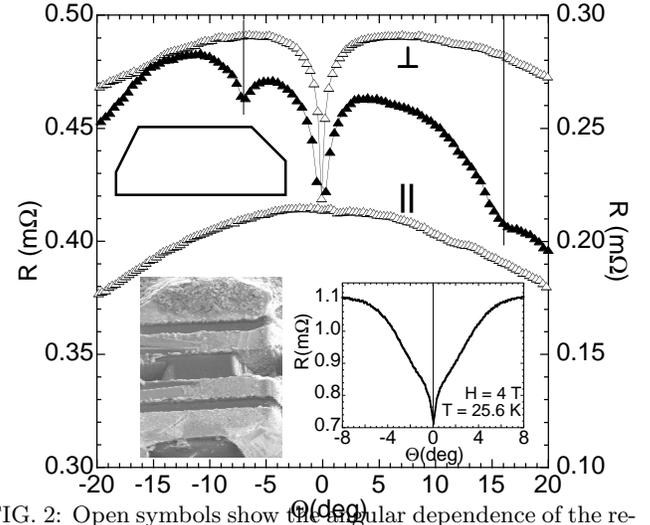}
\caption{Open symbols show the angular dependence of the resistance
when tilting the applied field across (``$\perp$'') and within
(``$\parallel$'') the side face.  The solid symbols show the
appearance of new dips after reshaping the crystal side surfaces with
an FIB. The data are taken at 2.5~T and 5~mA. The insets on the left
display a schematic cross section and an SEM image of the FIB-milled
crystal.  The inset on the right shows the cusp-like angular
dependence around the $ab$ plane.}
\label{Fig2}
\end{figure}

For exact field alignment with the $ab$ plane, the gradual resistive
onset becomes more pronounced with increasing fields, but is readily
suppressed by a tilt as small as $0.7^{\circ}$.  The strong angular
dependence is clearly seen in angular scans shown in Fig.~\ref{Fig2}
for both the $c$ axis (main panel) and the $ab$ plane (inset).
The angular dependence around
the $c$ axis depends on the direction in which the field is
tilted.  If the field is tilted across the vertical side faces (the
direction marked as ``$\perp$'' in the schematic inset of
Fig.~\ref{Fig1}a) the resistance increases rapidly in a cusp-like
fashion.  At high angles the resistance decreases again due to the
super-imposed, intrinsic, superconducting anisotropy.  If the field is
tilted in such a way that it stays parallel to the side surfaces
(marked as ``$\parallel$'' in Fig.\ref{Fig1}a) the resistance
remains at a low level. As a direct proof that the resistive behavior
shown in Fig.~\ref{Fig1} is associated with the sample
surfaces, we artificially modified the side faces of a crystal.
A trapezoidal cross-section was cut into the sample between the
voltage contacts (see insets of Fig.~\ref{Fig2}) using a
focused ion beam (FIB).
As a consequence, two new dips
appeared near $-7^{\circ}$ and $+16^{\circ}$ (solid symbols in
Fig.~\ref{Fig2}) corresponding to the orientation of the newly formed
surfaces. These results demonstrate that the
observed features in the transport properties indeed
originate from the crystal surfaces, and not from possible
planar, crystallographic defects that are aligned with the side
surfaces.

\begin{figure}
\includegraphics[width=1.0\linewidth]{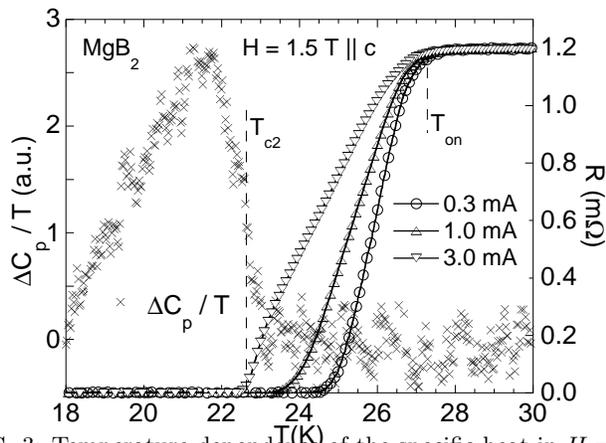}
\caption{Temperature dependence of the specific heat in $H=1.5$~T
$\parallel \mathbf{c}$ and of the resistance at various currents
measured on the same crystal.  The temperatures $T_{c2}$ and
$T_{\mathrm{on}}$ are indicated.}
\label{Fig3}
\end{figure}

It is important to note that the angular dependent, gradual,
superconducting onset at $T_{\mathrm{on}}$ in Figs.~\ref{Fig1} and
\ref{Fig2} occurs \emph{well above} the bulk upper critical field. 
Figure~\ref{Fig3} shows the specific heat signature of $T_{c2}$
together with the resistive transition of the \emph{same crystal}
measured at various currents in a field of $1.5$~T $\parallel
\mathbf{c}$.  The step in $\Delta C_{p}/T$, which corresponds to the
bulk superconducting transition, occurs about $5$~K below
$T_{\mathrm{on}}$ (for $H = 1.5$~T) but coincides well with the steep
drop to zero resistance seen at high enough currents.  This indicates
that the transport behavior is caused neither by a conventional
surface barrier effect \cite{MasuiCond}, nor by a transition in the
vortex system \cite{Eltsev02} for which the sample has to be in the
mixed state, i.e., below $H_{c2}$.  Instead, the normalconducting bulk
is covered by a superconducting sheath in places where the field is
aligned with the surface.

\begin{figure}
\includegraphics[width=1.0\linewidth]{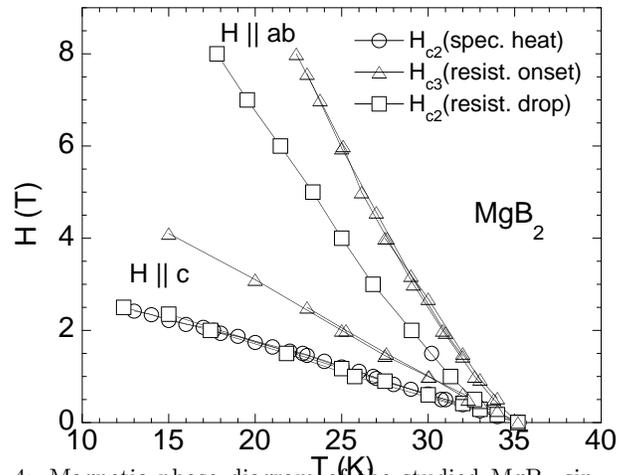}
\caption{Magnetic phase diagram of the studied MgB$_{2}$ single
crystals.  The bulk upper critical fields for the two main axes are
consistently obtained from specific heat and transport
measurements of the peak effect (which develops at the location of the
sharp resistive drop at high enough currents \cite{Welp03}).  Onsets
of transport nonlinearities and deviations from the normal state
resistivity are found at $\sim 1.6 H_{c2}^{c}$ for the $c$ axis and
$\sim 1.4 H_{c2}^{ab}$ for the basal plane.}
\label{Fig4}
\end{figure}

The field -- temperature ($H$--$T$) phase diagram resulting from the
data of Figs.~\ref{Fig1} and \ref{Fig3} and similar data is shown in
Fig.~\ref{Fig4}, and is in agreement with previous reports
\cite{Angst02,Lyard02,Welp03,Sologubenko02}.  Shown are the bulk upper
critical fields, as determined from specific heat measurements and
resistive drops, as well as the onsets of angular and current depend
transport.  We notice that the onsets are enhanced with respect to the
bulk $H_{c2}$ by a coefficient of about 1.6 for $\mathbf{H} \parallel
\mathbf{c}$ and 1.4 for $\mathbf{H} \parallel \mathbf{ab}$.  This
behavior is reminiscent of the development of a surface
superconducting state.  A superconducting surface sheath with a
thickness given by the Ginzburg--Landau coherence length, $\xi(T)$,
nucleates at an enhanced field of $H_{c3} = 1.69 H_{c2}$ when a
magnetic field is applied parallel to the flat (infinite) surface of
an isotropic superconductor \cite{Saint-James63}.  The exact value of
the enhancement factor may depend on clean-limit corrections
\cite{Hu69}, on the shape of the sample and its intrinsic anisotropy
\cite{Kogan02}, and on the surface quality \cite{Hart67}.

The results presented here can be accounted for in a model of surface
superconductivity.  It has been shown that the surface superconducting
state is rapidly suppressed in a cusp-like fashion when the applied
field has a normal component to the surface \cite{Hempstead64},
exactly the behavior observed when the field is tilted across the side
face (see Figs.~\ref{Fig1} and \ref{Fig2}).  Similarly, when the field
is tilted within the side faces the surface superconducting state is
not suppressed, and the only observed angular dependence arises from
the intrinsic anisotropy of MgB$_{2}$.  Furthermore, non-Ohmic
transport properties above the bulk upper critical field are expected
due to the presence of a surface critical current, $I_{\mathrm{sc}}$,
\cite{Hart67}.  If the applied current is smaller than
$I_{\mathrm{sc}}$ then the resistance will go to zero above $H_{c2}$
($T_{c2}$), see Fig.~\ref{Fig3}.  At higher current a finite, current
dependent resistance arises signaling current sharing between the
surface sheath and the normal core of the sample.  At $H_{c2}$ the
core of the sample goes superconducting, and the resistance drops to
zero.  We also note that (for macroscopic samples) the contribution of
the superconducting surface sheath has a negligible contribution to
the specific heat.

Figure~\ref{Fig1} and \ref{Fig3} suggest the surprising result that
the surface superconducting state is more pronounced and (as function
of angle) more robust for $\mathbf{H} \parallel \mathbf{c}$ than for
$\mathbf{H} \parallel \mathbf{ab}$, implying that surface
supercurrents on the narrow side faces (shaded in Fig.~\ref{Fig1}a)
are stronger than those on the wide top and bottom faces.  A similar
observation has recently been reported for NbSe$_{2}$ \cite{DAnna96}
for which indications of surface superconductivity were obtained for
$\mathbf{H} \parallel \mathbf{c}$ but not for $\mathbf{H} \parallel
\mathbf{ab}$.  Several factors may contribute to this behavior.  The
top and bottom surfaces appear very smooth indicating that the surface
critical currents are weak.  This is consistent with the broader
angular dependence for field around the $c$ axis and studies on PbTl
films and ribbons \cite{Hart67} that have shown that the angular
dependence is sharper for smooth surfaces.  In addition, due to the
anisotropy of the coherence length, the thickness of the current
carrying layer at the top and bottom faces is about four times smaller
than along the side faces.

\begin{figure}
\includegraphics[width=1.0\linewidth]{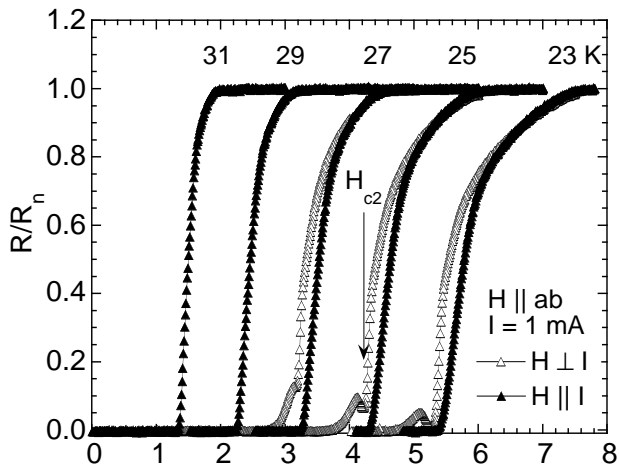}
\caption{Field dependence of the resistance for the zero Lorentz force
($\mathbf{H}\parallel \mathbf{I}$) and maximum Lorentz force
($\mathbf{H} \perp \mathbf{I}$) configurations.}
\label{Fig5}
\end{figure}

To complete the picture, we studied the influence of the angle between
the magnetic field and the current direction on the transport
properties, for $\mathbf{H}$ aligned with the basal plane. 
Figure~\ref{Fig5} shows the field dependence of the resistance for
$\mathbf{H} \perp \mathbf{I}$ and $\mathbf{H} \parallel \mathbf{I}$. 
A distinct orientational dependence is observed, reminiscent of the
Lorentz force effect on a vortex system.  In the parallel case the
transitions are featureless, while for $\mathbf{H} \perp \mathbf{I}$
the location of $H_{c2}$ is revealed by the already described, steep
drop of the resistance and the appearance of a peak effect at high
enough currents.  The angular dependence clearly persists to fields
well above $H_{c2}$.  In considing the surface superconductivity
description, the angular-dependence behavior arises
because the mutual orientation of $\mathbf{H}$ and $\mathbf{I}$
affects the distribution of the superconducting order parameter near
the surface \cite{Schmidt70}.  For $\mathbf{H} \perp \mathbf{I}$ the
superconducting order parameter is displaced with respect to the
surface \cite{Swartz67} whereas for $\mathbf{H} \parallel \mathbf{I}$
it acquires an additional phase analogous to the force free
configuration in thin, superconducting wires.  As a result there is an
intrinsic anisotropy of the superconducting phase stiffness and,
consequently, of the maximum (depairing) critical current
$I_{c}^{\perp}/I_{c}^{\parallel} = 0.6$ which can account for our
data.

Surface superconductivity as discussed here is a consequence of the
boundary conditions at the free surface of an otherwise unperturbed
sample.  In the case of MgB$_{2}$, ARPES experiments \cite{Uchiyama02}
show that the very existence of the surface induces modifications of
the electronic structure at Mg- and B-terminated surfaces.  Various
band structure calculations \cite{Kim01} indicate that these surface
electronic states could either locally enhance or suppress the
superconducting properties.  However, increasing the surface area by
pulverizing single crystals has not given any indication for enhanced
$T_{c}$ \cite{KarpinskiPrivate}.  In addition, the sample surfaces
could be modified due to exposure to oxygen and humidity. 
Although the value of $T_{c}$ has proven remarkably insensitive to
modest amounts of disorder \cite{Mazin02}, there could arise a shell
of enhanced (ÒdirtyÓ) upper critical field giving rise to the observed
transport behavior.  However, our experiments on the freshly cut
surfaces seem to rule out such a possibility.  Furthermore, the
observation of the general features shown here with enhancement
factors consistently in the range of 1.5 to 2 on a large number of
crystals from various sources indicates an intrinsic nature of the
surface effects.

In summary, we have shown that the surfaces of well-shaped MgB$_{2}$
single crystals possess locally enhanced superconducting properties in
magnetic fields aligned to the surfaces.  The surface
superconductivity is more pronounced for fields along the $c$ axis and
is found to display a field-induced anisotropy for fields in the basal
plane.  Further studies could address the possible significance of the
two-band structure for the exact mechanism behind the observations.

This work was supported through the Fulbright program and the
Sweden-America Foundation (A.R.), by the Ministry of Science and
Technology of Korea, and by the U.S. Department of Energy, Basic
Energy Sciences, under Contract No.~W-31-109-ENG-38.  Work with the
FIB was carried out in the Center for Microanalysis of Materials,
University of Illinois, which is partially supported by the U.S.
Department of Energy under grant DEFG02-91-ER45439.


\begin{thebibliography}{28}

\providecommand{\bibinfo}[2]{#2}
\providecommand{\eprint}[2][]{\url{#2}}

\bibitem{nagamatsu01}
J. Nagamatsu {\it et\,al.}, Nature (London) {\bf 410}, 63 (2001). For
extensive reviews see {\it Special Edition on MgB$_{2}$}, Physica
(Amsterdam) {\bf 385\/C} (2003).

\bibitem{Choi02}
H. J. Choi {\it et\,al.}, Nature (London) {\bf 418}, 758 (2002).

\bibitem{Liu01}
A. Y. Liu {\it et\,al.} Phys. Rev. Lett. {\bf 87}, 087005 (2001).

\bibitem{Bouquet01}
F. Bouquet {\it et\,al.}, Phys.  Rev.  Lett.  {\bf 87}, 047001 (2001);
P. Szabo {\it et\,al.}, {\it ibid} {\bf 87}, 137005 (2001); F.
Giubileo {\it et\,al.}, {\it ibid} {\bf 87} 177008 (2001); H. Schmidt
{\it et\,al.}, {\it ibid} {\bf 88}, 127002 (2002); M. Iavarone {\it
et\,al.}, {\it ibid} {\bf 89}, 187002 (2002).

\bibitem{Angst02}
M. Angst {\it et\,al.}, Phys.  Rev.  Lett.  {\bf 88}, 167004 (2002);
S. L. BudÕko {\it et\,al.}, Phys.  Rev.  B {\bf 64}, 180506 (2001); M.
Zehetmayer {\it et\,al.}, {\it ibid} {\bf 66}, 052505 (2002); Y.
Machida {\it et\,al.}, {\it ibid} {\bf 67}, 094507 (2003).

\bibitem{Lyard02}
L. Lyard {\it et\,al.}, Phys. Rev. B {\bf 66}, 180502 (2002).

\bibitem{Welp03}
U. Welp {\it et\,al.}, Phys. Rev. B {\bf 67}, 012505 (2003).

\bibitem{Sologubenko02}
A. V. Sologubenko {\it et\,al.}, Phys. Rev. B {\bf 65}, 180505 (2002).

\bibitem{Eskildsen02}
M. Eskildsen {\it et\,al.}, Phys. Rev. Lett. {\bf 89}, 187003 (2002).

\bibitem{Bouquet02}
F. Bouquet {\it et\,al.}, Phys. Rev. Lett. {\bf 89}, 257001 (2002)

\bibitem{Pradhan02}
A. K. Pradhan {\it et\,al.}, Phys.  Rev.  B {\bf 65}, 144513 (2002);
Yu.  Eltsev, Physica (Amsterdam) {\bf 385\/C}, 162 (2003).

\bibitem{Eltsev02}
Yu.  Eltsev {\it et\,al.}, Physica (Amsterdam) {\bf 378--381\/C}, 61
(2002); Yu.  Eltsev {\it et\,al.}, Phys.  Rev.  B {\bf 65}, 140501
(2002).

\bibitem{MasuiCond}
T. Masui {\it et\,al.}, cond-mat/0210358.

\bibitem{Jung02}
C. U. Jung {\it et\,al.}, Phys. Rev. B {\bf 66}, 184519 (2002).

\bibitem{Kim02}
K. H. Kim {\it et\,al.}, Phys. Rev. B {\bf 65}, 100510 (2002).

\bibitem{Saint-James63}
D. Saint-James and P. G. de Gennes, Phys.  Lett.  {\bf 7}, 306 (1963); A.
A. Abrikosov, Sov.  Phys.  JETP {\bf 20}, 480 (1965); H. J. Fink,
Phys.  Rev.  Lett.  {\bf 14}, 309 (1965).

\bibitem{Hu69}
C.-R. Hu and V. Korenman, Phys.  Rev.  {\bf 178}, 684 (1969); Phys. 
Rev.  B {\bf 6}, 1 (1972).

\bibitem{Kogan02}
V. G. Kogan {\it et\,al.}, Phys. Rev. B {\bf 65}, 094514 (2002).

\bibitem{Hempstead64}
C. Hempstead and Y. Kim, Phys.  Rev.  Lett.  {\bf 12}, 145 (1964); W. J.
Tomasch and A. S. Joseph, Phys.  Rev.  Lett.  {\bf 12}, 148 (1964); R. S.
Thompson, Sov.  Phys.  JETP {\bf 42}, 1144 (1976).

\bibitem{Hart67}
H. R. Hart, Jr., and P. S. Swartz, Phys. Rev. {\bf 156}, 403 (1967).

\bibitem{Kulik69}
I. O. Kulik, Soviet Phys. JETP {\bf 28}, 461 (1969).

\bibitem{DAnna96}
G. D{'}Anna {\it et\,al.}, Phys. Rev. B {\bf 54}, 6583 (1996).

\bibitem{Schmidt70}
H. Schmidt and H. J. Mikeska, J. Low Temp.  Phys.  {\bf 3}, 123 (1970);
A. E. Koshelev, to be published.

\bibitem{Swartz67}
This effect predicts rectification as seen, for example, in P. S.
Swartz and H. R. Hart, Jr., Phys.  Rev {\bf 156}, 412 (1967).  We
could not observe any such effects here, however, possibly due to the
contact arrangement.

\bibitem{Uchiyama02}
H. Uchiyama {\it et\,al.}, Phys. Rev. Lett. {\bf 88}, 157002 (2002);
S. Souma {\it et\,al.}, Nature {\bf 423}, 65 (2003). 

\bibitem{Kim01}
I. G. Kim {\it et\,al.}, Phys.  Rev.  B {\bf 64}, 020508 (2001); V. M.
Silkin {\it et\,al.}, {\it ibid} 172512; E. Bascones and F. Guinea,
{\it ibid} 214508; Z. Li {\it et\,al.}, Phys.  Rev.  B {\bf 65},
100507 (2002); V. D. Servedio {\it et\,al.}, Phys.  Rev.  B {\bf 66},
140502 (2002); G. Profeta {\it et\,al.}, {\it ibid} 184517.

\bibitem{KarpinskiPrivate}
J. Karpinski, private communication.

\bibitem{Mazin02}
I. I. Mazin {\it et\,al.}, Phys. Rev. Lett. {\bf 89}, 107002 (2002).

\end{thebibliography}
\end{document}